\documentclass{research4cacm}
\usepackage{mathtools}
\newcommand\myeq{\stackrel{\mathclap{\normalfont\mbox{def}}}{=}}
\newtheorem{definition}{Definition}

\begin{document}
%

\title{Data Science: Challenges and Directions
%
}
%
%
%
%
%
\numberofauthors{2} 
%
\author{
%
%
\alignauthor
Longbing Cao\\
}

\maketitle

\textbf{Key insights:} 
\begin{itemize}
\item Data science problems are complex systems, thus requiring systematic thinking, methodologies and approaches for understanding and managing challenges.
\item A systematic picture is given about various low-level complexities, intelligence, and research prospects in the data science discipline.
\item Data science challenges may cause violations of assumptions taken in existing theories and systems, hence suggesting significant breakthroughs in developing data science theories and simulating human-like intelligence.
\end{itemize}

While data science has emerged as a contentious new scientific field, enormous debates and discussions have been made on it why we need data science and what makes it as a science. However, only a limited number of discussions are about intrinsic complexities and intelligence embedded in data science problems, and the gaps and opportunities for disciplinary directions. 

After a comprehensive review \cite{cao16-2,Cao16pe,Diggle15,Donoho15,Huber11,Lazer14,Mattmann13} of hundreds of literature that directly incorporates data science in their scopes, we make the following observations of the big data buzz and data science debate:
\begin{itemize}
\item Very comprehensive discussion has taken place, not only within data-related or data-focused disciplines and domains, such as statistics, computing and informatics, but also in the non-traditional data-related fields and areas such as social science and management. Data science has thus emerged as an inter- and cross-disciplinary new field.
\item Although many discussions and publications are available, most (probably more than 95\%) essentially concern existing concepts and topics discussed in statistics, data mining, machine learning and broad data analytics. This demonstrates how data science has developed and been transformed from existing core disciplines, in particular, statistics, computing and informatics, etc.
\item While data science as a term has been increasingly used in publications and media, it seems that most authors have done this to make the work look `sexier'. The abuse, misuse and over-use of  the term ``data science'' is ubiquitous, and essentially contributes to the buzz and hype. Myths and pitfalls are everywhere \cite{Cao16np}.
\item While specific challenges have been discussed \cite{mckinsey11,Jagadish14}, very limited articles are available that address the low-level complexities and problematic nature of data science, or contribute deep insights about the intrinsic challenges, directions and opportunities of data science as a new field.
\end{itemize} 

Our experience and literature review also confirm that data science enables new opportunities for new scientific research: i.e., ``what I can do now but could not do before'' (e.g., processing large scale data), ``what I could do before but does not work now'' (e.g., those methods that assume data objects are IID), ``problems that have not been solved well before are becoming even more complex'' (e.g., quantifying complex behavioral data), and better innovation: i.e., ``what I could not do better before'' (e.g., deep learning). 

As data science focuses on a more comprehensive and systematic view \cite{cao16-2,Cao16pe}, this article will particularly draw on the viewpoint that data science problems are complex systems \cite{Mitchel2011,Metasynthetic15} and data science tasks are to transform data to knowledge and intelligence for decision making. Hence, the discussions focus on complexities, knowledge and intelligence hidden in complex data science problems, and the opportunities for disciplinary development of data science from a complex system perspective. 


\section{What Is Data Science}
\label{sec:ds}
The concept of ``data science'' was originally proposed in the statistics and mathematics community \cite{Tukey62,Tukey77}, at which time it essentially concerned data analysis. 
Today, the art of data science \cite{Matsudaira15} goes beyond specific areas like data mining and machine learning, and the argument that data science is the next-generation of statistics \cite{Cleveland01,Donoho15,Huber11}. Data science is becoming a very rich concept which carries the vision and responsibilities of an independent scientific field that is systematic and inter-disciplinary.

So what is data science? 

\begin{definition}[Data Science$^2$]\label{def:dsdiscipline}
Data science is a new trans-disciplinary field that builds on and synthesizes a number of relevant disciplines and bodies of knowledge, such as statistics, informatics, computing, communication, management and sociology, to study data and its domain following a data science thinking.
\end{definition}


As an example, which is shown in Fig. \ref{fig:disciplinary-ds-definition}, a \textit{discipline-based data science formula} is given below: 
\begin{eqnarray}
data~science \myeq \{statistics ~\cap ~informatics~ \cap ~computing~ \nonumber \\ \cap ~communication~ \cap ~sociology~ \cap ~management~ | ~data~ \cap \nonumber \\  ~domain~ \cap ~thinking\} 
\end{eqnarray}
where ``$|$'' means ``conditional on.'' 


\begin{figure}
\centerline{\includegraphics[scale=0.5]{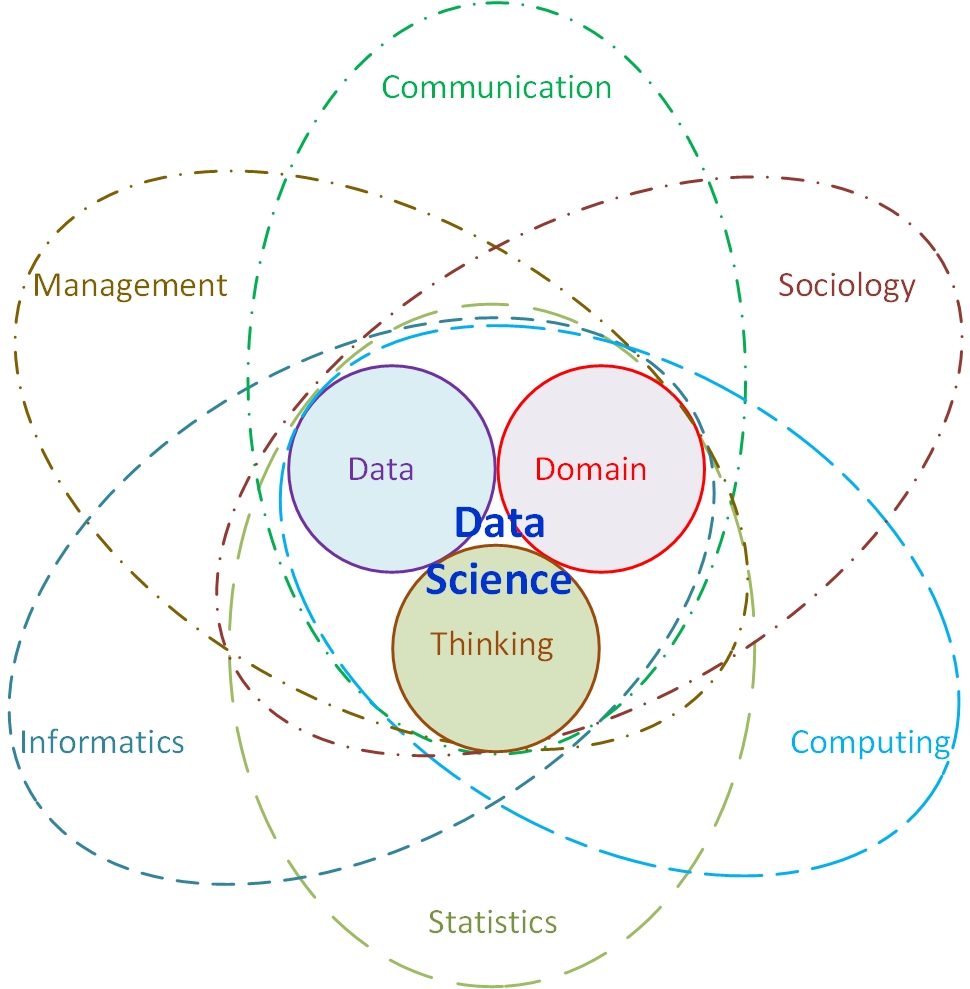}}
\caption{Trans-disciplinary data science.}
\label{fig:disciplinary-ds-definition}
\end{figure}

\section{X-Complexities in Data Science}
\label{subsec:complexity}

A core objective of data science innovation is to effectively explore the sophisticated and comprehensive complexities \cite{Mitchel2011} trapped in data, business, data science tasks, and problem-solving processes and systems, which form a complex system \cite{Metasynthetic15}. Here \textit{complexity} refers to sophisticated characteristics in data science systems. We treat a data science problem as a complex system, in which comprehensive system complexities are embedded, named \textit{X-complexities}, in terms of data (characteristics), behavior, domain, societal aspects, environment (context), learning (process and system), and complex deliverables.

\emph{Data complexity} is reflected in terms of sophisticated data circumstances and characteristics, such as largeness of scale, high dimensionality, extreme imbalance, online and real-time engagement, cross-media applications, mixed sources, strong dynamics, high frequency, uncertainty, noise mixed with valuable data, unclear structures, unclear hierarchy, heterogeneous or unclear distribution, strong sparsity, and unclear availability of specific data. A very important issue concerns the \emph{complex relations} hidden in data and business, which form a key component of data characteristics, and are critical in properly understanding the hidden driving forces in data and business. Complex relations may consist of comprehensive couplings \cite{noniid14} that may not be describable by existing association, correlation, dependence and causality theories and systems. Learning mixed explicit and implicit couplings, structural relations, non-structural relations, semantic relations, hierarchical and vertical relations, relation evolution and reasoning are critical and challenging. Some of the above mentioned data complexities may propose new perspectives that could not be done or done better before. For example, in traditional large survey of sensor data, statisticians design questions and sample participants to be surveyed. This has shown to be ineffective, represented by issues such as low overall response rate and many questions unanswered. This way is even more problematic for designing representative, categorized, and personalized Web-scale survey, which could be better done by data-driven discovery of who to be surveyed, what questions to be answered, and how cost-effective the survey would be. 

\emph{Behavior complexity} becomes increasingly visible in understanding what actually takes place in business, as behaviors carry the semantics and processes of behavioral objects and subjects in the \emph{physical world} that are often ignored or largely simplified in the transformed \emph{data world} after the physical-to-data conversion undertaken by the existing data management systems. Behavior complexities are embodied in such aspects as coupled individual and group behaviors, behavior networking, collective behaviors, behavior divergence and convergence, non-occurring behaviors, behavior network evolution, group behavior reasoning, the insights, impact, utility and effect of behaviors, the recovery of what actually happened, happens or will happen in the physical world from the highly deformed information collected in the data world, and the emergence of behavior intelligence. However, quantifying and analyzing complex behaviors has not been explored well.

\emph{Domain complexity} has become increasingly recognized \cite{dddm10} as a critical aspect for deeply and genuinely discovering data characteristics, value and actionable insights. Domain complexities are reflected in such aspects as domain factors, domain processes, norms, policies, qualitative to quantitative domain knowledge, expert knowledge, hypotheses, meta-knowledge, the involvement of and interaction with domain experts and professionals, multiple and cross-domain interactions, experience acquisition, human-machine synthesis, roles and leadership in the domain. However, the related work mainly focuses on involving domain knowledge.

\emph{Social complexity} is embedded in business and data, and its existence is inevitable in data and business understanding. It may be embodied in such aspects as social networking, community emergence, social dynamics, impact evolution, social conventions, social contexts, social cognition, social intelligence, social media networking, group formation and evolution, group interactions and collaborations, economic and cultural factors, social norms, emotion, sentiment and opinion spreading and influence processes, and social issues including security, privacy, trust, risk and accountability in social contexts. Enormous interdisciplinary opportunities appear when social science meets data science.

\emph{Environment complexity} plays an important role in complex data and business understanding. It is reflected in environmental factors, relevant contexts, context dynamics, adaptive engagement of contexts, complex contextual interactions between environment and data systems, significant changes in environment and their impact on data systems, and variations and uncertainty in the interactions between data and environment. Such aspects have been concerned in open complex systems \cite{Qian93} but not yet in data science. If ignored, a model suitable for one domain may produce misleading outcomes for another, as often seen in recommender systems. 
 
\emph{Learning (process) complexity} has to be properly addressed to achieve the goal of data analytics. Typical challenges include developing effective methodologies, common task frameworks and learning paradigms to handle various aspects of data, domain, behavioral, social and environmental complexity. For example, there are additional challenges in learning multiple sources and inputs, parallel and distributed inputs, heterogeneous inputs, and dynamics in real time; supporting on-the-fly active and adaptive learning, as well as ensemble learning while considering the relations and interactions between ensembles; supporting hierarchical learning across significantly different inputs; enabling combined learning across multiple learning objectives, sources, feature sets, analytical methods, frameworks and outcomes; and learning non-IID data mixing coupling relationships with heterogeneity \cite{noniid14}. 

Other matters include the appropriate design of experiments and mechanisms. Inappropriate learning could result in misleading or harmful outcomes, e.g., a classifier works for balanced data would mistakenly classify biased and sparse cases for wrongly anomaly detection. 

\emph{Deliverable complexity} becomes an issue when actionable insights \cite{dddm10} are focused in data science. This necessitates the identification and evaluation of  the outcomes that satisfy technical significance and have high business value from both objective and subjective perspectives. The challenges are also embedded in designing the appropriate evaluation, presentation, visualization, refinement and prescription of learning outcomes and deliverables to satisfy diversified business needs, stakeholders, and decision purposes. 
In general, deliverables to business are expected to be easy to understand and interpretable from the non-professional perspective, disclosing and presenting insights that directly inform and enable decision-making actions and possibly having a transformative effect on business processes and problem-solving. 

\section{X-Intelligence in Data Science}
\label{subsec:ui}
Data science is an intelligence science. The nature of data science is the drive to achieve a successful transformation from data to knowledge, intelligence and wisdom \cite{Rowley07}. During this process, comprehensive intelligence \cite{Metasynthetic15}, here termed ``X-intelligence'', 
is often involved in a complex data science problem, from data to domain, organizational, social and human aspects, and the representation and synthesis of them. 
Here \textit{X-intelligence} refers to comprehensive information that informs or supports the deeper, more structured and organized comprehension, representation and problem-solving of underlying complexities and challenges.
Below, we discuss the X-intelligence associated with the different aspects of complexity discussed in Section \ref{subsec:complexity}.

\emph{Data intelligence} highlights the interesting information and stories about the formation of business problems or driving forces and their reflection in the corresponding data. Intelligence hidden in data is obtained through understanding specific data characteristics and complexities. Apart from the usual focus on exploring the complexities in data structures, distribution, quantity, speed, and quality issues from the individual data object perspective, the focus in data science is on the intelligence hidden in the unknown space D in Figure \ref{fig:k-unkown}. For example, in addition to existing protocols for cancer treatments, what are new ways that are informed by historical treatments and patient feedback?
The level of data intelligence is dependent on how much and to what extent we can completely understand and capture data characteristics and complexities.

\emph{Behavior intelligence} is discovered by understanding the activities, processes, dynamics and impact of individual and group actors who are the data quantifiers, owners and users in the physical world. This requires to 
bridge the gaps between the data world and the physical world by connecting what happened, happens and will happen to formation and dynamics of the real world problem, and to discover behavior insights through developing \textit{behavior informatics} \cite{Cao10-1}. 
For example, in online shopping websites, one challenge is to recognize whether and how some ratings and comments are made by robots; similarly, in social media, detecting robot-triggered comments in billions of daily transactions is extremely challenging. Constructing behavior sequences and interactions with other accounts in a time period and then differentiating abnormal behaviors may be a useful way to understand the difference between  proactive and subjective human activities and reactive behavior patterns of robots.

\emph{Domain intelligence} emerges from properly involving relevant domain factors, knowledge and meta-knowledge, and other domain-specific resources that not only wrap a problem and its target data but also assist in problem understanding and the development of problem-solving solutions. Involving qualitative and quantitative domain intelligence can inform and enable a deep understanding of domain complexities and their critical roles in discovering unknown knowledge and actionable insights. For example, to design effective high-frequency trading strategies, we have to involve the orderbook and microstructure of limit market into modeling. 

\emph{Human intelligence} plays a critical or centric role in complex data science systems, through the explicit or direct involvement of human empirical knowledge, belief, intention, expectation, run-time supervision, evaluation and expert groups. It also concerns the implicit or indirect involvement of  human intelligence as imaginary thinking, emotional intelligence, inspiration, brainstorming, reasoning inputs and embodied cognition such as convergent thinking through interactions with other members in the process of data science problem-solving. For example, as thinking is crucial for data science, data scientists may have to apply subjective factors, qualitative reasoning, and critical imagination. 

\emph{Network intelligence} emerges from both Web intelligence and broad-based networking and connected (especially social media networks and mobile services) activities and resources such as information and resource distribution, linkages between distributed objects, hidden communities and groups, information and resources from networks, and, in particular, the Web, distributed and cloud infrastructure and computing facilities, information retrieval, searching, and structuralization from distributed repositories and the environment. The information and facilities from the networks surrounding a target business problem either serve as the problem constituents or contribute to useful information for complex data science problem solving. A relevant example is the crowdsourcing-based open source system development and algorithm design.

\emph{Organizational intelligence} emerges from 
the understanding and involvement of organizational goals, actors and roles, as well as organizational structures, behaviors, evolution and dynamics, governance, regulation, convention, process and workflow in data science systems. For example, the cost-effectiveness of enterprise analytics and functioning of data science team rely on the proper engagement of organizational intelligence.   

\emph{Social intelligence} consists of human social intelligence and animated intelligence which emerge from social complexities discussed in the above Section. 
Human social intelligence is related to such aspects as social interactions, group goals and intentions, social cognition, emotional intelligence, consensus construction, and group decision-making. Social intelligence is often associated with social network intelligence and collective interactions, as well as the business rules, law, trust and reputation for governing the emergence and use of social intelligence. Typical areas in which social intelligence is focused include social networks and social media, in which data-driven social complexities are understood, such as social influence modeling and understanding community formation and evolution in virtual online society.

\emph{Environmental intelligence} refers to the intelligence hidden in the environment of a data science problem. This can be specified in terms of and broadly connected to the underlying domain, organizational, social, human and network intelligence. Data science systems are open, with the interactions between the converted data world and the transformed physical world as the broad environment. Examples are context-aware analytics which involves contextual factors and evolving interactions and changes between data and context, such as in infinite dynamic relation modeling.

\section{Known-to-Unknown Data to Decision Transformation}
\label{subsec:k-unkown}

We view a complex data science problem-solving journey as a cognitive progression from known to unknown complexities in order to transform data to knowledge, intelligence and insights for decision and action taking by inventing and applying respective capabilities. 
Here \textit{knowledge} represents the form of processed information in terms of an information mixture, procedural actions, or propositional rules, \textit{insight} refers to the genuine and deep understanding of intrinsic complexities and working mechanisms in data and its corresponding physical worlds. 

\begin{figure}
\centerline{\includegraphics[width=0.5\textwidth]{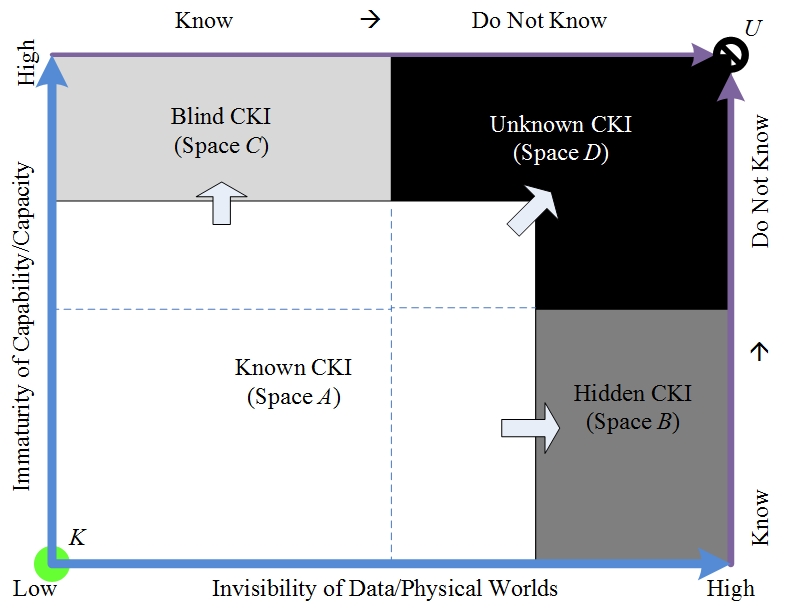}}
\caption{Data science: Known-to-unknown discovery and progression.}
\label{fig:k-unkown}
\end{figure}

Figure \ref{fig:k-unkown} illustrates the data science progression that is to reduce the \textit{immaturity of capabilities and capacity} (\textit{y}-axis) to better understand the \textit{invisibility of complexity, knowledge and intelligence (CKI) in data/physical worlds} (\textit{x}-axis) from the 100\% known state \textit{K} to the 100\% unknown state \textit{U}. 
According to the status and levels of data/physical world visibility and capability/capacity maturity, our recognition about a data science problem can be categorized into four statuses. 
 
Space A represents the \textit{known space}: i.e., \textit{I (my mature capability/capacity) know what I know (about visible world)}. This is similar to the ability of sighted people being able to recognize an elephant by seeing the whole animal, whereas non-sighted people might only get to identify part of the animal by touch. The knowledge in the visible data is known to people with mature capability/capacity (the level of capability/capacity maturity is sufficient to understand the level of data/physical world invisibility). This space refers to a well-understood status in the community. Examples include profiling and descriptive analysis, which apply existing models to data deemed to follow certain assumptions. 

Space B represents the \textit{hidden space}: i.e., \textit{I know what I do not know (about the invisible world)}. For some people or disciplines, although their certain capability/capacity is mature, CKI is hidden to (cannot be addressed by) the current level of maturity of the capability/capacity in data science; more advanced capabilities/capacity are required. An example is the existing IID models such as k-means and KNN which cannot handle non-IID data. 
Another situation is the Space C representing the \textit{blind space}: i.e., \textit{I (my immature capability) do not know what I know (about the world)}. Although the CKI is visible to some people or disciplines, their capability/capacity is also mature, but they do not match well; the immaturity makes the world blind to them. An example is a well-established social scientist starts to handle data science problems.  

Lastly, Space D represents the \textit{unknown space}: i.e., \textit{I do not know what I do not know}, CKI in the invisible world is unknown as a result of immature capability. This is the area that data science future research and discovery are focused. 
When the world invisibility increases, the capability immaturity also grows. In the fast-evolving big data world, the CKI invisibility increases, resulting in increasingly larger unknown space.

\begin{figure}
\centerline{\includegraphics[scale=0.5]{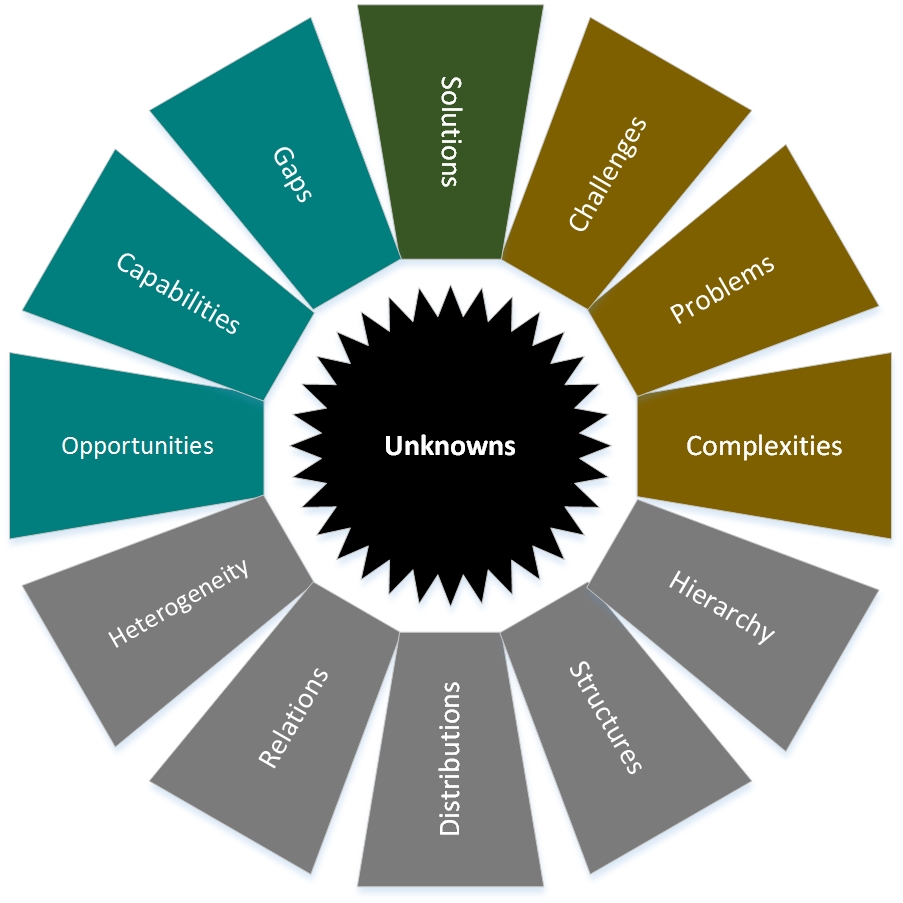}}
\caption{Data science: The unknown world.}
\label{fig:unknown-world}
\end{figure}

The stage ``we do not know what we do not know'' can be explained in terms of various unknown perspectives and scenarios. As shown in Fig. \ref{fig:unknown-world}, the unknown world presents unknownness in terms of (1) problems, challenges, and complexities; (2) hierarchy, structures, distributions, relations, and heterogeneities, (3) capabilities, opportunities, and gaps, and (4) solutions. 

\section{The Disciplinary Directions}
\label{subsec:advanalytics}

In this section, a conceptual landscape is discussed, followed by two significant data science issues: non-IID data learning and human-like intelligence revolution.

\subsection{Data Science Landscape}
\label{subsec:landscape}
\begin{figure}
\centerline{\includegraphics[width=0.5\textwidth]{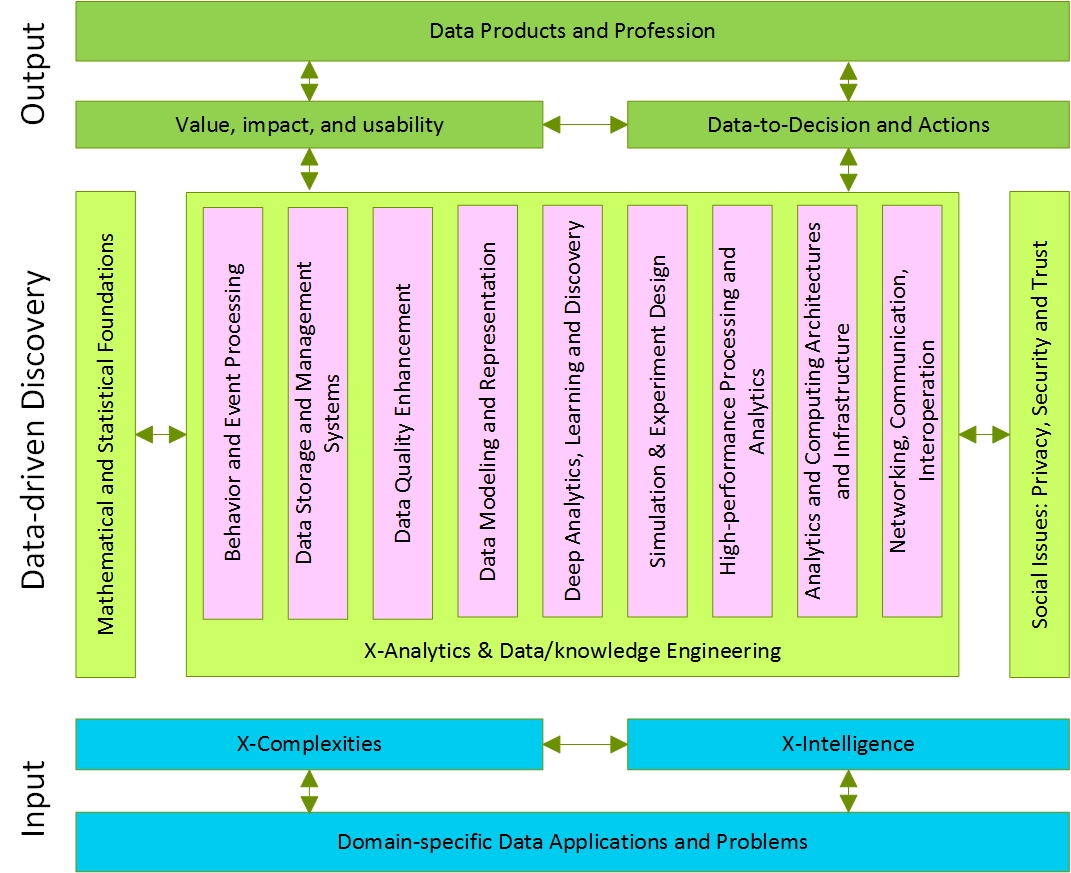}}
\caption{Data science conceptual landscape.}
\label{fig:seven}
\end{figure}

The X-complexity and X-intelligence in complex data science systems, and the increasing gaps between world invisibility and capability/capacity immaturity bring new research challenges which form data science a new discipline. Figure \ref{fig:seven} illustrates the conceptual landscape of data science and its major research issues by taking an interdisciplinary, complex system-based, and hierarchical view. 

As shown in Figure \ref{fig:seven}, the data science landscape consists of three layers: the \textit{data input} including domain-specific data applications and systems, X-complexity and X-intelligence in the data and business, the \textit{data-driven discovery} consisting of a collection of discovery tasks and challenges, and the \textit{data output} composed of various results and outcomes.

Research challenges and opportunities emerge from all three layers, which are categorized in terms of five major areas that cannot be managed well by existing methodologies, theories and systems.
\begin{itemize}
\item \textit{Data/business understanding challenges}: This is to identify, specify, represent and quantify the X-complexities and X-intelligence that cannot be managed well by existing theories and techniques but nevertheless exist and are embedded in a domain-specific data and business problem. Examples are to understand in what forms, at what level, and to what extent the respective complexities and intelligence interact and integrate with each other, and to devise effective methodologies and technologies for incorporating them into data science tasks and processes.  
\item \textit{Mathematical and statistical foundation challenges}: This is to discover and explore whether, how and why existing theoretical foundations are insufficient, missing, or problematic in disclosing, describing, representing, and capturing the above complexities and intelligence and obtaining actionable insights. Existing theories may need to be extended or substantially redeveloped so as to cater for the complexities in complex data and business, for example, supporting multiple, heterogeneous and large scale hypothesis testing and survey design, learning inconsistency, change and uncertainty across multiple sources of data, enabling large scale fine-grained personalized predictions, supporting non-IID data analysis, and creating scalable, transparent, flexible, interpretable, personalized and parameter-free modeling.
\item \textit{Data/knowledge engineering  and X-analytics challenges}: This is to develop domain-specific analytic theories, tools and systems that are not available in the body of knowledge, to represent, discover, implement and manage the relevant and resultant data, knowledge and intelligence, and to support the corresponding data and analytics engineering. Examples are autonomous and automated analytical software that can automate the process, and self-monitor, self-diagnose and self-adapt to data characteristics and domain-specific context, and learning algorithms that can recognize data complexities and self-train the corresponding optimal models customized for the data. 
\item \textit{Quality and social issues challenges}: This is to identify, specify and respect social issues related to the domain-specific data and business understanding and data science processes, including processing and protecting privacy, security and trust and enabling social issues-based data science tasks, which have not previously been handled well. Examples are privacy-preserving analytical algorithms, and benchmarking the trustfulness of analytical outcomes. 
\item \textit{Data value, impact and utility challenges}: This is to identify, specify, quantify and evaluate the value, impact and utility associated with domain-specific data that cannot be addressed by existing theories and systems, from technical, business, subjective and objective perspectives. Examples are the development of measurement for actionability, utility and values of data.
\item \textit{Data-to-decision and action-taking challenges}: This is to develop decision-support theories and systems to enable data-driven decision generation, insight-to-decision transformation, and decision-making action generation, incorporating prescriptive actions and strategies into production, and data-driven decision management and governance which cannot be managed by existing technologies and systems. Examples include tools for transforming analytical findings to decision-making actions or intervention strategies. 
\end{itemize}

Since data/knowledge engineering and X-analytics play the keystone role in data science, we discuss specific research issues that have not been addressed satisfactorily.  
\textit{Data quality enhancement} is fundamental, which handles both existing data quality issues such as noise, uncertainty, missing values and imbalance, which may present to a very different extent and level due to the significantly increasing scale and extent of the complexity, and new data quality issues emerging in the big data and Internet-based data/business environment, such as cross-organizational, cross-media, cross-cultural, and cross-economic mechanism data science problems. 

\textit{Data modeling, learning and mining} faces the challenge of modeling, learning, analyzing and mining data that is embedded with X-complexities and X-intelligence. For example, \textit{deep analytics} is essential to discover unknown knowledge and intelligence hidden in the unknown space in Figure \ref{fig:k-unkown} that cannot be handled by existing latent learning and descriptive and predictive analytics; another opportunity is to integrate data-driven and model-based problem-solving, which balances common learning models and frameworks and domain-specific data complexities and intelligence-driven evidence learning.

X-complexity and X-intelligence propose new challenges to \textit{simulation and experimental design}. Issues include how to simulate the respective complexities and intelligence, working mechanisms, processes, dynamics and evolution in data and business, and how to design experiments and explore the effect and impact if certain data-driven decisions and actions are undertaken in the business.

Big data analytics requires \textit{high-performance processing and analytics}, which needs to support large scale, real-time, online, high frequency, Internet-based cross-organizational data processing and analytics while balancing local and global resource involvement and objectives. This may generate new distributed, parallel and high-performance infrastructure, batch, array, memory, disk and cloud-based processing and storage, data structure and management systems, and data to knowledge management. 

Complex data science tasks also pose challenges to \textit{analytics and computing architectures and infrastructure}, e.g., how to enable the above tasks and processes by inventing efficient analytics and computing architectures and infrastructure based on memory, disk, cloud and Internet-based resources and facilities. Another important matter is how to support the \textit{networking, communication and interoperation} between different data science roles in a distributed data science team and during the whole-of-cycle of data science problem-solving. This requires the distributed cooperative management of projects, data, goals, tasks, models, outcomes, workflows, task scheduling, version control, reporting and governance. 

The exploration of the above issues in data science and analytics requires systematic and interdisciplinary approaches. This may require synergy between many related research areas, including data representation, preparation and preprocessing, distributed systems and information processing, parallel computing, high performance computing, cloud computing, data management, fuzzy systems, neural networks, evolutionary computation, system architecture, enterprise infrastructure, network and communication, interoperation, data modeling, data analytics, data mining, machine learning, cloud computing, service computing, simulation, evaluation, business process management, industry transformation, project management, enterprise information systems, privacy processing, information security, trust and reputation, business intelligence, business value, business impact modeling, and the utility of data and services. This is owing to the need of 
addressing critical complexities in complex data science problems that cannot be addressed by singular disciplinary efforts. For instance, new data structures and detection algorithms are required to handle high frequency real-time risk analytics issues in extremely large online businesses, such as online shopping and cross-market trading.

\subsection{Assumption Violations in Data Science}
\label{subsec:noniidness}

Big data is complex, which owns certain X-complexities discussed in Section \ref{subsec:complexity}, including complex coupling relationships and/or mixed distributions, formats, types and variables, and unstructured and weakly structured data. Such complex data has proposed significant challenges to many existing mathematical, statistical, and analytical methods which have been built on certain assumptions, owing to the fact that these assumptions are violated in big data. Many models and methods come up with certain assumptions. When these assumptions do not hold, the modeling outcomes may be inaccurate, distorting, misleading, or even faulty. In addition to general scenarios, such as whether data violates the assumptions of normal distribution, t-test, and linear regression, assumption check applies to broad aspects, including independence, normality, linearity, variance, randomization, and measurement that apply to population data and analysis. 

There is not much fundamental work undertaking in the relevant communities on detecting and verifying such validations, and even less work on inventing new theories and tools to manage and circumvent the assumption violations. One of such violations highlighted here is the independent and identically distributed (IID) assumption, because big/complex data (referring to objects, values, attributes, and other aspects \cite{noniid14}) is essentially non-IID, whereas most of existing analytical methods are IID  \cite{noniid14}. 

\begin{figure}
\centerline{\includegraphics[width=0.35\textwidth]{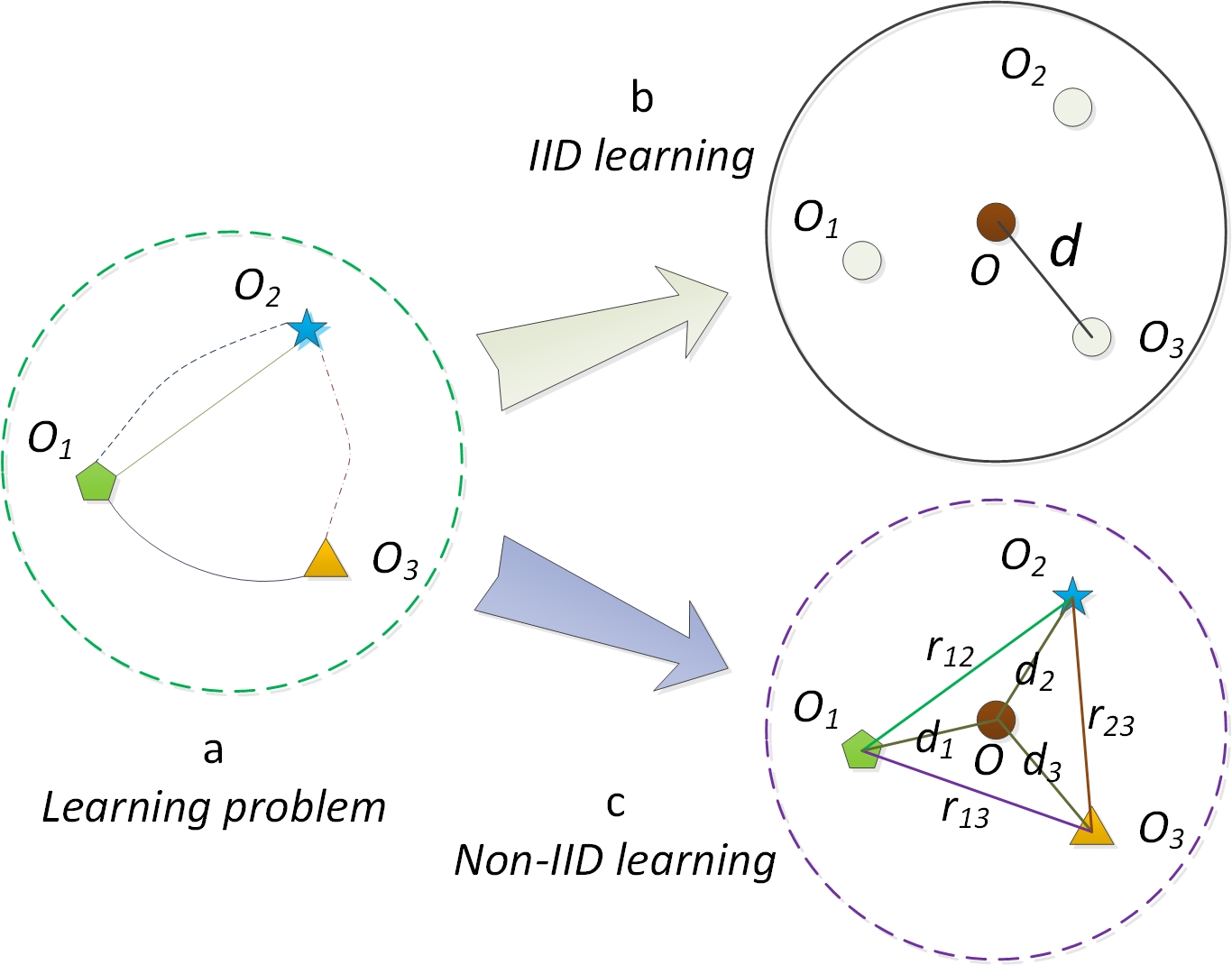}}
\caption{IIDness vs. non-IIDness in data science problems.}
\label{fig:eight}
\end{figure}

In a non-IID data problem (see Figure \ref{fig:eight}(a)), \emph{non-IIDness} (see Figure \ref{fig:eight}(c)) refers to any \textit{couplings} (both well-explored relationships such as co-occurrence, neighborhood, dependency, linkage, correlation, and causality, and poorly-explored and ill-structured ones such as sophisticated cultural and religious connections and influence) and \textit{heterogeneity}, which exist within and between two or more aspects, such as entity, entity class, entity property (variable), process, fact and state of affairs, or other types of entities or properties (such as learners and learned results) appearing or produced prior to, during and after a target process (such as a learning task). By contrast, \textit{IIDness} ignores or simplifies them, as shown in Figure \ref{fig:eight}(b). 

Learning visible and especially invisible non-IIDness is fundamental for a deep understanding of data with weak and/or unclear structures, distributions, relationships, and semantics. In many cases, locally visible but globally invisible (or vice versa) non-IIDness are presented in a range of forms, structures, and layers and on diverse entities. Often, individual learners cannot tell the whole story due to their inability to identify such complex non-IIDness. Effectively learning the widespread, various, visible and invisible non-IIDness is thus crucial for obtaining the truth and a complete picture of the underlying problem.

We frequently only focus on explicit non-IIDness, which is visible to us and easy to learn. Typically, work in the hybridization of multiple methods and the combination of multiple sources of data into a big table for analysis fall into this category. Computing non-IIDness refers to understanding, formalizing and quantifying the non-IID aspects, entities, interactions, layers, forms and strength. This includes extracting, discovering and estimating the interactions and heterogeneity between learning components, including the method, objective, task, level, dimension, process, measure and outcome, especially when the learning involves multiples of one of the above components, such as multi-methods or multi-tasks. We are  concerned about understanding non-IIDness at a range of levels from values, attributes, objects, methods and measures to processing outcomes (such as mined patterns). Such non-IIDness is both comprehensive and complex. 

Below, we illustrate the main prospects of inventing new and effective data science theories and tools for \textit{non-IIDness learning} or \textit{non-IID data learning} \cite{noniid14}. We examine how to address the non-IID data characteristics (note, not just about IID objects) in terms of new feature analysis by considering feature relations and distributions, new learning theories, algorithms and models for analytics, and new metrics for similarity measurement and evaluation.
\begin{itemize}
\item Deep understanding of non-IID data characteristics: This is to identify, specify and quantify non-IID data characteristics, factors, aspects, forms, types, levels of non-IIDness in data and business, and identify the difference between what can be captured by existing data/business understanding technologies and systems and what is left out.
\item New and effective non-IID feature analysis and construction: This is to invent new theories and tools for the analysis of feature relationships by considering non-IIDness within and between features and objects, and developing new theories and algorithms for selecting, mining and constructing features. 
\item New non-IID learning theories, algorithms and models: This is to create new theories, algorithms and models for analyzing, learning, and mining non-IID data by considering value-to-object couplings and heterogeneity.
\item New non-IID similarity and evaluation metrics: This is to develop new similarity and dissimilarity learning methods and metrics, as well as evaluation metrics that consider non-IIDness in data and business.
\end{itemize}

More broadly, many existing data-oriented theories, designs, mechanisms, systems and tools may need to be re-invented when non-IIDness is taken into consideration. In addition to non-IIDness learning for data mining, machine learning and general data analytics, this involves well-established bodies of knowledge, including mathematical and statistical foundations, descriptive analytics theories and tools, data management theories and systems, information retrieval theories and tools, multi-media analysis, and X-analytics. 

\subsection{Understanding Data Characteristics and Complexities} 
\label{subsec:understandingdatacharacteristics}

To address critical issues like assumption violations, we believe data characteristics and data complexities determine their values, complexities in data modeling, and quality of data-driven discovery. 

\textit{Data characteristics} refer to the profile and complexities of data (in general, a data set), which can be described in terms of many aspects of data such as distribution, structure, hierarchy, dimension, granularity, heterogeneity, and uncertainty.

Understanding data characteristics is concerned with the following fundamental challenges and directions \cite{Cao16pe}:
\begin{itemize}
\item What data characteristics are, namely, how to define data characteristics?
\item How to represent and model data characteristics, namely, how to quantify the different aspects of data characteristics?
\item How to conduct data characteristics-driven data understanding, analysis, learning and management? and
\item how to evaluate the quality of data understanding, analysis, learning and management in terms of data characteristics?  
\end{itemize}

Unfortunately, very limited research outcomes and systematic theories and tools are available. Answering these questions represent some grant challenges in data science. 

\subsection{Data Brain and Human-like Machine Intelligence}
\label{subsec:hmi}

It is often debated whether machines could replace humans \cite{Suchma06}. While it may not be possible to build \textit{data brain} and \textit{intelligent thinking machines} that have identical abilities to humans, big data analytics and data science are driving the revolution from logical thinking-centered machine intelligence to imaginary thinking-oriented ``non-traditional''  machine intelligence. This may be partially evidenced by the Google AlphaGo success of beating Lee Sedol \cite{Deepmind}, the Facebook emotion experiment \cite{Kramer14}, but none of these actually exhibits human-like imaginary thinking. This transformation in machine thinking, such as by implementing \emph{data science thinking} \cite{Cao16pe}, if it is able to mimic the above human intelligence well, may reform machine intelligence and significantly or even fundamentally change the current man-machine role and segmentation of responsibilities.    

Data science and big data analytics present new opportunities to promote the human-like machine intelligence revolution in terms of building several new mechanisms in machines or upgrading existing machine intelligence.  First, a critical capability of humans is to be \textit{curious}, starting from childhood. We want to know what, how and why. \textit{Curiosity} connects other cognitive activities, in particular, imagination, reasoning, aggregation, creativity and enthusiasm, which then often produce new ideas, observations, concepts, knowledge, and decisions. During this process, human intelligence is upgraded. Accordingly, a critical task is to enable machines to generate and retain curiosity through learning inquisitively from data and generating curiosity in data.  

Second, human \textit{imaginary thinking} differentiates humans from machines, which have sense-effect working mechanisms. Human \textit{imagination} is creative, evolving, and even uncertain, which cannot be generated by following patterns and pre-defined sense-effect mechanisms. This requires data analytical algorithms and systems to simulate human imagination processes and mechanisms, before creative machines are available. Existing knowledge representation, reasoning and aggregation and computational logics, reasoning and logic thinking incorporated into machines do not support curiosity and imagination, and machines are not \textit{creative}. Correspondingly, the existing computer theories, operating systems, system architectures and infrastructures, computing languages, and data management need to be fundamentally reformed. One way to do this is to simulate, learn, reason and synthesize from data and engage other intelligence in a non-predefined and patternable way, in contrast to existing simulation, learning and computation which are largely predefined by default.    
  
Further, the exploration of X-intelligence in complex data problems requires learning the \textit{micro-meso-societal level} of hierarchical complexities and intelligence. A major future direction is to progress toward imaginary thinking (i.e., non-logical thinking) and new (networked) mechanisms for invisibility learning, knowledge creation, complex reasoning, and consensus building through connecting heterogeneous and relevant data and intelligence. 

Lastly, \textit{data-analytical thinking}, a core part of \emph{data science thinking} \cite{Cao16pe}, needs to be built into data products and data professionals. \textit{Data-analytical thinking} is not only explicit, descriptive and predictive, but also implicit and prescriptive. It mimics human thinking by involving and synthesizing comprehensive data, information, knowledge and intelligence through various cognitive processing methods and processes.

A data-driven human-like machine may develop abilities and capabilities to simulate the working mechanism of the human brain, in particular, human imaginary thinking and processing; learn and absorb societal and human intelligence hidden in data and business during the problem-solving process; understand unstructured and mixed structured data and intelligence, to extract these structures, and to convert the unstructured data and intelligence to structured representation and models; understand qualitative problems and factors, and quantify qualitative factors and problems to form quantitative representations and models; observe, measure and learn human behaviors and societal activities, and to evaluate and select preferred behaviors and activities to undertake; synthesize collective intelligence to solve problems that cannot be handled by individuals; generate knowledge of knowledge (abstraction and summarization) and derive new knowledge based on implicit and networked connections in existing data, knowledge and processing; ask questions actively and be motivated by online learning and inspiration from certain learning procedure, objects and groups; be capable of creating and discovering new knowledge adaptively and online; gain insights and optimal solutions to address grand problems on a web-scale or global scale through  experiments on massively possible hypotheses, scenarios and trials; and provide personalized and evolving services and decisions based on an intrinsic understanding of personal characteristics, behaviors, needs, emotion and changes in circumstance.

\section{Methodologies for Complex Data Science Systems}
\label{subsec:methodology}

The complexities discussed in Section \ref{subsec:complexity} and the X-intelligence discussed in Section \ref{subsec:ui} in major data science and analytics tasks render a complex data project equivalent to an open complex intelligent system \cite{Metasynthetic15}. 
Building such complex intelligent systems require effective methodologies to understand, specify, quantify and manipulate the X-complexities and X-intelligence. 

The use of X-intelligence may take one of the following two paths: \emph{single intelligence engagement} or \emph{multi-aspect intelligence engagement}. An example of single intelligence engagement is the involvement of domain knowledge in data mining and the consideration of user preferences in recommender systems. This applies to simple data science problem solving and systems. In general, multi-aspect X-intelligence exists in complex data science problems. 

\begin{figure}
\centerline{\includegraphics[width=0.5\textwidth]{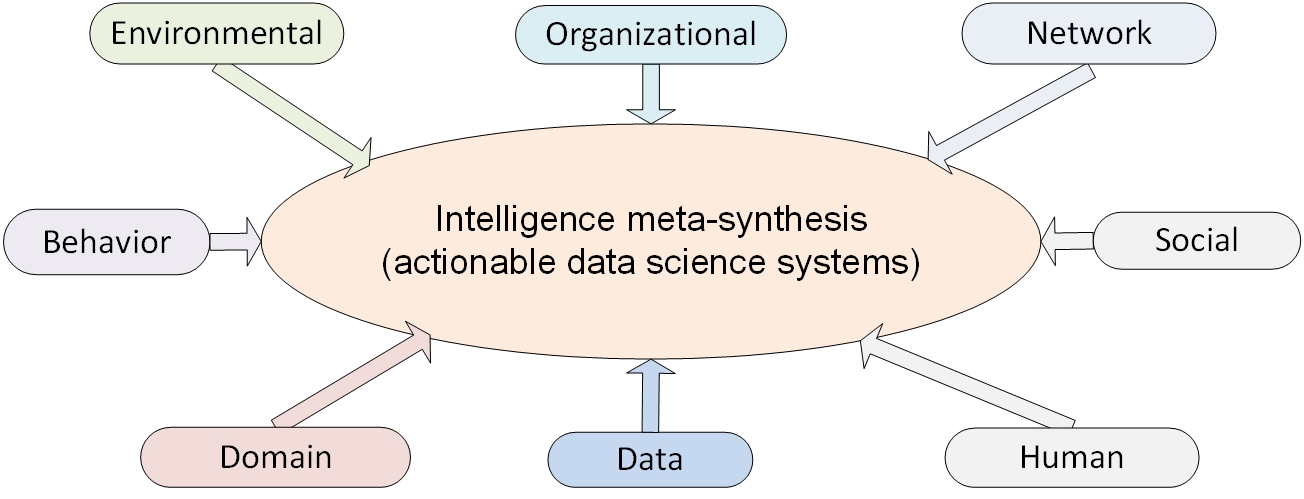}}
\caption{Synthesizing X-intelligence in data science.}
\label{fig:six}
\end{figure}

As shown in Figure \ref{fig:six}, the performance of a data science problem-solving system is highly dependent on the effective recognition, acquisition, representation and integration of relevant intelligence and indicative factors from human, domain, organization and society, network and web perspectives. For this, new methodologies and techniques need to be developed. The theory of \textit{metasynthetic engineering} \cite{Qian93,Metasynthetic15} and the approach to the \textit{integration of ubiquitous intelligence} may provide useful methodologies and techniques for synthesizing X-intelligence in complex data and analytics. 

From a high level perspective, the principle of intelligence meta-synthesis \cite{Qian93,Metasynthetic15} is to involve, synthesize and use ubiquitous intelligence in the complex data and environment to discover actionable knowledge and insights \cite{dddm10}. 
The process for intelligence meta-synthesis to solve complex data science problems involves a complex system engineering, in which several aspects of complexities and intelligence are often embedded in the data, environment and problem-solving process. 
Simply using the \textit{reductionism} methodology \cite{Metasynthetic15} for data and knowledge exploration may not work well. This is because the problem may not initially be clear, certain, specific and quantitative, thus it cannot be effectively decomposed and analyzed. Further, the analysis of the whole does not equal the sum of the analysis of the parts (this is the common challenge of complex systems) \cite{Qian93}. 

Accordingly, the theories of \textit{system complexities} and the corresponding complex system methodologies \emph{systematism} (or \emph{systematology}, combination of reductionism with holism) \cite{Qian93,Metasynthetic15} - may then be applicable for the analysis, design and evaluation of complex data science problems. 

When a data science problem involves large scale objects, multiple levels of sub-tasks or objects, multiple sources and types of data objects from online, business, mobile or social networks, complicated contexts, human involvement and domain constraints, it presents the characteristics of an open complex system \cite{Qian93,Metasynthetic15}. It is likely to present typical \textit{system complexities}, including openness, large or giant scale, hierarchy, human involvement, societal characteristics, dynamic characteristics, uncertainty and imprecision \cite{Qian93,Mitchel2011,Metasynthetic15}.
  
Typically, a big data analytical task satisfies most if not all of the above system complexities. To address such problems, one possibly effective methodology is the \textit{qualitative-to-quantitative metasynthesis} \cite{Qian93,Metasynthetic15}, which was initially proposed to guide the engineering of open complex giant systems \cite{Qian93}. This qualitative-to-quantitative metasynthesis supports the exploration of open complex systems by engaging various intelligences. In implementing this methodology for engineering open complex intelligent systems, the metasynthetic computing and engineering (MCE) approach \cite{Metasynthetic15} provides a systematic computing and engineering guide and a suite of tools to build the framework, processes, analysis and design tools for engineering and computing complex systems.  

\begin{figure}
\centerline{\includegraphics[width=0.5\textwidth]{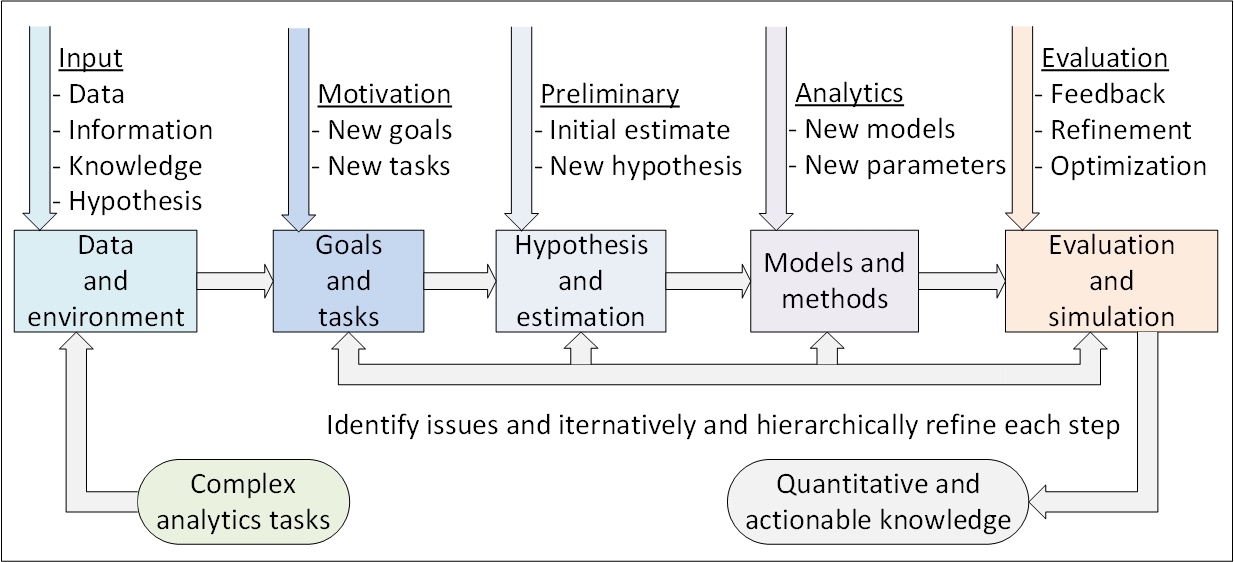}}
\caption{Complex data science problems: qualitative-to-quantitative X-intelligence metasynthesis.}
\label{fig:mec}
\end{figure}

Figure \ref{fig:mec} illustrates the process of applying the qualitative-to-quantitative metasynthesis methodology to address a complex analytical problem. For a complex analytics task, the MCE approach supports an iterative and hierarchical problem-solving process, starting by incorporating the corresponding input, including data, information, domain knowledge, initial hypothesis and underlying environmental factors. Motivations are set for analytics goals and tasks to be explored on the data and environment. With preliminary observations obtained from domain and experience, hypotheses and estimations are identified and verified, which guide development of the modeling and analytics method. Findings are then evaluated and simulated, which are fed back to the corresponding procedures for refinement, optimization and adjustment, towards the achievement of new goals, tasks, hypotheses, models and parameters, when appropriate. Following these iterative and hierarchical explorations of qualitative-to-quantitative intelligence, quantitative and actionable knowledge is identified and delivered to address data complexities and analytical goals. 

As an example, \textit{domain driven data mining} \cite{dddm10} integrates diversified intelligence for complex knowledge discovery problems. It advocates a comprehensive process of interaction and integration between multiple kinds of intelligence, as well as the encouragement of intelligence emergence toward delivering actionable knowledge. This goal is achieved by way of properly understanding data characteristics as the most important task in analytics; acquiring and representing unstructured, ill-structured and uncertain domain/human knowledge; supporting the dynamic involvement of business experts and their knowledge/intelligence in the analytics process; acquiring and representing expert thinking such as imaginary thinking and creative thinking in group heuristic discussions during data understanding and analytics; acquiring and representing group/collective interaction behaviors and their impact; and building infrastructure that supports the involvement and synthesis of ubiquitous intelligence.

\section{Conclusion}
\label{sec:concl}

The low-level complexities and intelligence in complex data science problems determine the gaps between the world invisibility and our capability immaturity. This requires a disciplinary effort and the development of complex data science thinking and methodology from  complex system perspective. 

The possible disciplinary revolution of data science creates unique opportunities for breakthrough research, cutting-edge technological innovation, and significant new data business. If parallels are drawn between the evolution of the Internet and the evolution of data science, the future and impact of data science may be unpredictable.


\bibliographystyle{abbrv} 
\bibliography{cacm-ds.20}  

\textbf{Longbing Cao} (longbing.cao@gmail.com) is a professor at the Advanced Analytics Institute in the University of Technology Sydney, Australia.

\balancecolumns
\end{document}